\documentclass[12pt,a4paper]{article}

\usepackage{amsmath}
\usepackage{amsfonts}
\usepackage{amssymb}

\def\pa{\partial}

\def\s{\sigma}

\def\p{\varphi}
\def\k{\kappa}

\def\d{\delta}
\def\r{\rho}
\def\h{\hat}

\def\l{\label}
\def\a{\alpha} 
\def\bt{\beta}

\def\m{\mu}
\def\n{\nu}
\def\be{\begin{equation}}
\def\ee{\end{equation}}
\def\ba{\begin{eqnarray}}
\def\ea{\end{eqnarray}}

\def\h{\hat}

\def\f{\frac}
\def\G{\Gamma}\vspace{4cm}

\def\e{\epsilon}

\begin{document}

\centerline{\large\bf Connection between two forms of extra dimensional metrics revisited }
\vspace{2cm}
\centerline{\bf Mikhail Z. Iofa \footnote{ e-mail:
iofa@theory.sinp.msu.ru}} \centerline{Skobeltsyn Institute of Nuclear
Physics}
\centerline{Moscow State University}
\centerline{Moscow 119992, Russia}
\vspace{1 cm}

\begin{abstract}
5D cosmological model with 3-brane with matter is considered. 
The brane divides bulk in two AdS half spaces. 
Geometry of the model can be described by two types of coordinates: in the first setting the 
 metric is static and the brane is moving in the bulk,  in the second approach the metric is time-dependent  
and the brane is located at a fixed position in the bulk. Coordinate transformation connecting two 
coordinate systems is constructed. 
\end{abstract}

Recently cosmological models with extra dimensions have attracted a lot of attention \cite{maart}).
  We consider a 5D model with infinite extra dimension with the 
brane with matter embedded in the bulk. The brane divides the bulk in two 
AdS spaces at both sides of the brane. 
The model can be considered in 
two alternative approaches. In the first approach the bulk is static, 
and the brane is moving in the bulk \cite{kraus,col,cham,bir,bow}. In the second approach metric of 
the model is time-dependent, and the brane is located at a fixed 
position in extra dimension \cite{BDL1, BDL2}. 
Coordinate  transformation between two settings is not
trivial and was discussed previously \cite{bow,muk} and recently \cite{wu}.
Below, omitting various generalizations of the model, we construct transformation between two
pictures in a simple and straightforward way.

The energy-momentum tensor on the brane is
$$
\tau_\m^\n =diag\{-\h{\r}-\h{\s}, \h{p}-\h{\s},\h{p}-\h{\s},\h{p}-\h{\s}
\},
$$
where $\h{\r}$ and $\h{p}$ are the energy and momentum densities of
matter on the brane, $\h{\s}$ is the tension of the brane.

It is convenient to define the normalized energy density and tension of the
brane
$$
\r =\f{\k^2\h{\r}}{6}, \qquad \s =\f{\k^2\h{\s}}{6},
$$
where $\h{\r}$ and $\h{\s}$ have dimension $[mass]^4$ and $\r$ and $\s$
have dimension $[mass]^1$.
Parameter $\m$ having dimension $[mass]$ is defined via  the 5D
cosmological constant $\Lambda$ as $\m^2 =\k^2 |\Lambda|/6$, where $\k^2 =8\pi/M^3$ is
the 5D gravitational coupling constant,
and $M$ is the 5D gravitational scale.

In  approach with the static metric, the metric is 
\be
\l{1}
ds^2 = -f(R) dT^2 +\f{d R^2}{f(R)} +\m^2 R^2 dx^a dx_a \equiv g_{MN}dx^M dx^N,
\ee
where
$$
f(R)= \m^2 R^2 - \f{P}{R^2}
.$$
$R$ and $T$ have dimension $[mass]^{-1}$.

In a parametric form trajectory of the brane is defined through the proper time $t$ on the
brane as
$R= R_b (t),\,\,T= T_b (t)$, where $ -f(R_b )\dot{T}_b^2
+f^{-1}(R_b )\dot{R}_b^2 =-1 $. It follows that
\be
\l{2}
\dot{T}=\xi \frac{\sqrt{f(R_b )+\dot{R}_b^2}}{f(R_b )},
\ee
where $\xi =\pm 1$.
 The metric on the brane is $ds^2 =-dt^2 +\m^2  R_b^2 (t) dx^a dx_a$.

Trajectory of the brane with fixed 3D coordinates is determined by solving the junction
conditions for the metric on the brane.
The junction conditions on the brane are \cite{cham,col,BDL1,BDL2,maeda}
\be
\l{3}
[h^k_i\nabla_k n_j] = \tau_{ij} -\f{1}{3}\tau h_{ij}
.\ee
Here $h_{ij}= g_{ij} -n_i n_j$ is the induced metric on the brane, $v^i =\pm (\dot{T}_b 
,\dot{R}_b
,0)$ and $n_i =\pm (-\dot{R}_b ,\dot{T}_b ,0)$ are velocity and normal vector to the brane 
$\tau_{ij} = (\r +p)v_i v_j + p h_{ij}$,
 [X] denotes the
difference of expressions calculated at the opposite sides of the brane.

From the spatial components ($a,b=1,2,3$) of the junction conditions
follows the generalized Friedmann equation
\cite{kraus,col,cham,BDL1,BDL2,maeda,maart}
\be
\l{4}
\left(\f{\dot{R_b}}{R_b}\right)^2 =-\m^2 +(\r +\s )^2 +\f{ P}{R_b^4}
.\ee

In the second approach  we consider a class of non-static metrics 
\cite{BDL1,BDL2}
\be
\l{5}
ds^2 =- n^2  (y,t) dt^2 +  a^2 (y,t) dx^a dx_a +dy^2 .
\ee
The brane is located at a fixed position in
the extra coordinate, which we set $y=0$.
The function
$n(y,t)$ is normalized by
condition $ n(0,t) =1$.
For simplicity we consider a spatially flat brane.
Here
\ba
\l{6}
&{}&a^2 (y,t) =\f{a^2 (0,t)}{4}\left[e^{2\m |y|}
\left(\left(\f{\r +\s}{\m}-1 \right)^2 +\f{\r_w}{\m}\right)+
e^{-2\m |y|} \left(\left(\f{\r +\s}{\m}+1 \right)^2
+\f{\r_w}{\m}\right)\right.\\\nonumber
&{}& \left.-2\left(\left(\f{\r +\s}{\m}\right)^2 -1
+\f{\r_w}{\m}\right)\right]
,\ea
and $ n(y,t) =\dot{a}(y,t)/\dot{a}(0,t)$.
Restriction of the metric (\ref{5}) to the brane is 
 $ds^2 =-dt^2 +a^2 (t) dx^a dx_a$.
The function $a(t)=  a(0,t)$ satisfies  generalized Friedmann
equation  \cite{BDL1,BDL2}
\be
\l{7}
H^2 (t) = -\m^2 +(\r +\s )^2 +\m \r_w (t)
,\ee
where
$$
H(t) =\f{\dot{a}(0,t)}{a(0,t)},\qquad  \r_w (t) =\f{ \r_{w0}}{a^4 (0,t)}.
$$
are the Hubble function and the Weyl radiation term.
In derivation of (\ref{6}) $\r_{w0}$ appears as an integration constant \cite{BDL2}.
Eqs (\ref{4}) and
 (\ref{7}) have the same form.
 $ a^2 (0,t)$ can be identified with $ \m^2  R_b^2 (t)$, and
the term $\r_w (t) = \r_{w0}/a^4 (t)$ can be identified with the term
$P/\m R_b^4 (t)$.
%%%%%%%%%%%%%%%%%%%%%%%%%%

In the picture with the static metric  we consider the geodesic equations 
starting at the brane world sheet
\ba
\l{8}
&{}&\f{d^2 T}{dy^2}+2\G^T_{TR}\f{dT}{dy}\f{dR}{dy}=0
\\\l{1g3}
&{}&\f{d^2 x^a}{dy^2}+ 2\G^a_{bR}\f{dx^b}{dy}\f{dR}{dy}=0
\\\l{a7}
&{}&\f{d^2 R}{dy^2}+ \G^R_{RR}\left(\f{dR}{dy}\right)^2 +
\G^R_{TT}\left(\f{dT}{dy}\right)^2
+\G^R_{ab}\f{dx^a}{dy}\f{dx^b}{dy}=0,
\ea
where $y$ is the affine parameter and
 the Christoffel symbols are
$$
\G^T_{TR}=\f{f'}{2f},\quad \G^R_{RR}=-\f{f'}{2f},\quad
\G^R_{TT}=\f{1}{2}ff' ,\quad \G^R_{ab} =-\eta_{ab}f\m^2 R,\quad
\G^a_{Rb}=\f{\d^a_b}{R}
.$$
Here $(T,\, R)\equiv (T^\pm ,\, R^\pm )$ are coordinates in the AdS spaces
at the opposite sides of the brane.
Integrating the geodesic equations, one  obtains
\be
\l{9}
\f{dT^\pm}{dy}=\f{E^\pm}{f(R)},\qquad \f{dx^a}{dy}=\f{C^a}{\m^2 R^2},
\qquad
\left(\f{dR^\pm}{dy}\right)^2 =f(R) ({C^{R\,\pm}})^2 +{E^{\pm}}^2 -\f{{C^a}^2 f}{\m^2
R^2}
,\ee
where $ (E^\pm,\,C^{a} , \,C^{R\,\pm} ) $ are integration parameters.
$(dT/dy,\,dR/dy,\,dx^a /dy)$ are the components of the tangent vectors to
geodesics which we normalize to unity.
Imposing the normalization condition
\be
\l{n}
\f{dx^M}{dy}\f{dx^M}{dy}g_{MN} =1, \qquad M,N=T,R,a
\ee
we obtain that $({C^{R\,\pm}})^2 =1$.
We consider solutions of the geodesic equations even in $y$:
$T^+ (y)=T^- (-y),\,\, R^+ (y)=R^- (-y)$ . 

The hypersurface $(T,\,R,\,x^a =0)$ is foliated by  geodesics that intersect the
trajectory of the
brane $(T_b (t),\,R_b (t))$ and at the intersection point  are orthogonal to the world sheet of 
the brane  (cf \cite{muk}).
 Parameter $t$ is constant along the geodesics orthogonal to
the brane world sheet and can be considered as a 
parameter labeling the geodesics.
Setting $C^a =0$, have
\be
\l{10}
\f{\pa T^{\pm}(y,t)}{\pa y}=\f{E\e (y)}{f(R)},
\qquad
\f{\pa R^{\pm}(y,t)}{\pa y} =\a\e (y)\left(f(R) +{E}^2 \right)^{1/2}
,\ee
where $\a =\pm 1$, and $E^+ =E^- =E$.

 The normalized velocity vector to the trajectory of the brane and
the normal vectors to the hyperplane containing trajectory of the brane $(T_b (t),\,R_b (t))$ are
\be
\l{11}
v^i_b =(\dot{T}_b ,\,\dot{R}_b) =\left(\xi\sqrt{f(R_b )+\dot{R}_b^2}/{f(R_b
)},\,\dot{R}_b)\right)\qquad
n^{i\pm}_b =\eta\e(y) \left(\f{\dot{R}_b}{f(R_b )} ,\,\xi\sqrt{f(R_b )+\dot{R}_b^2} \right)
,\ee
where  $\eta =\pm 1$.
From (\ref{10}) we obtain the tangent vector to a geodesic
\be
\l{12}
u^{i\pm} =\left(\f{E\e (y)}{f(R)},\,\,
\a\e (y)\sqrt{f(R)+E^2}\right)
,\ee
By construction, the tangent vector to a geodesic at the intersection point with
trajectory of the brane is (anti)parallel to the normal to the trajectory of the brane,
\be
\l{13}
u^i\,|_{y=0} || n^i_b
.\ee
From this condition it follows that
\be
\l{14}
E=\eta \dot{R}_b ,\qquad \a=\xi\eta
\ee

Denoting $Q^i = (R,\,T)$ and $u^i = \pa Q^i/\pa y$, and introducing $a_j =u^i g_{ij}$, normalization 
condition 
(\ref{n}) with $C^a =0$ can be written as
$$
a_j \f{\pa Q^j (y,t)}{\pa y}\bigg|_t =1
.$$
From this equation we obtain
\be
\l{15}
a_j =\f{\pa y}{\pa Q^j} +\p(y,t) \f{\pa t}{\pa Q^j},
\ee
where $\p(y,t)$ is an arbitrary function
\footnote{This remark is due to I. Tyutin and B. Voronov.}.
Although in a general case I cannot prove that $\p=0$, in a special case discussed below
 it is possible to have $\p=0$.

If $\p =0$, (\ref{15}) yields $\pa y/\pa Q^i =g_{ij} \pa Q^j /\pa y$, or
\be
\l{18}
\f{\pa y}{\pa R}= \f{1}{f(R)}\f{\pa R(y,t)}{\pa y},\qquad \f{\pa y}{\pa T}= -f(R)\f{\pa T(y,t)}{\pa y}
\ee
From the equality 
$$
dy=\f{\pa y}{\pa R}\left(\f{\pa R}{\pa y}dy +\f{\pa R}{\pa t}dt\right)+
\f{\pa y}{\pa T}\left(\f{\pa T}{\pa y}dy +\f{\pa T}{\pa t}dt\right) 
$$
it follows that
\be
\l{16}
\f{\pa y}{\pa R}\f{\pa R}{\pa t}+\f{\pa y}{\pa T}\f{\pa T}{\pa t}=0
.\ee
Using (\ref{18}), this relation is  written in a form
\be
\l{116}
\f{\pa T(y,t)/\pa t}{\pa R(y,t)/\pa t}=\f{1}{f^2 (R)}\f{\pa R(y,t)/\pa y}{\pa T(y,t)/\pa y}.
\ee
Now it is possible to transform the metric (\ref{1}) to the form (\ref{5}).
The metric (\ref{1}) is written as
\be
\l{17}
ds^2 =dy^2\left(-f(R){T'}^2 +\f{{R'}^2}{f(R)}\right) +2dy\, 
dt\left(-f(R){T'}\dot{T}+\f{R'\dot{R}}{f(R)}\right)+
dt^2 \left(-f(R){\dot{T}}^2 +\f{{\dot{R}}^2}{f(R)}\right) 
+\m^2 R^2 dx^adx_a
\ee
Using (\ref{10}) we find that the coefficient at $dy^2$ is 1, using (\ref{10}),(\ref{18}),(\ref{116}) 
that the 
coefficient at $dydt$ is zero, and at $dy^2$ is $\dot{R}^2 /\dot{R_b}^2 =n^2 (y,t)$.

Let us consider a special case $P=0$ in (\ref{1}).
Integrating the second equation (\ref{10}), we obtain
\be
\l{19}
R^\pm (y,t)=
R_b (t)\cosh\m y +\a\sqrt{\f{\dot{R}_b^2}{\m^2} +R_b^2}\,\sinh\m |y|
.\ee
 Using the Friedmann equation  with $P=0$
\be
\l{20}
H^2 =\dot{R}_b^2 /R^2_b =\r^2 +2\m\r
\ee
 and introducing $\bt =sign (R_b (t))$, we rewrite (\ref{19}) as
\be
\l{20}
R^\pm (y,t) =R_b (t)\left(\cosh\m y +\a\bt \left (1 +\f {\r}{\m}\right)\sinh \m |y|\right)
.\ee
For correspondence with (\ref{6}) (with $P=0$)
$$
a^2 (y,t)=\f{a^2 (0,t)}{4}\left[e^{2\m|y|}\left(\f{\r}{\m}\right)^2
+e^{-2\m|y|}\left(\f{\r}{\m}+2\right)^2 -2\f{\r}{\m}\left(\f{\r}{\m}+2\right)\right]
$$
we set
$$
\a\bt =-1
$$
 and (omitting $(\pm))$ obtain
\be
\l{21}
R(y,t) =\f{R_b (t)}{2}\left[e^{-\m |y|}\left(\f{\r}{\m}+2\right)-e^{\m |y|}\f{\r}{\m}\right].
\ee
Also we have
\be
\l{23}
\f{d R(y,t)}{dy}=-\e (y)\f{\m R_b}{2}\left[e^{-\m |y|}\left(\f{\r}{\m}+2\right)+e^{\m
|y|}\f{\r}{\m}\right]
.\ee
Introducing $y_0$, such that
\be
\l{24}
e^{\m y_0}=\left(\frac{\r}{\r +2\m}\right)^{1/2}
\ee
we express $R(y,t)$ and $R' (y,t)$ as
\be
\l{25}
R(y,t)=-\f{\dot{R}_b (t)}{\m}\sinh (\m |y| +\m y_0 ),\qquad R' (y,t)=-\e (y) \dot{R}_b (t)\cosh (\m 
|y|+\m y_0 )
\ee
Integrating the first  Eq. (\ref{10}) for $T(y,t)$ with $R(y,t)$ (\ref{21}), we obtain
\be
\l{26}
T^\pm (y,t) =-\f{1}{\m E}\f{\cosh (\m |y| +\m y_0 )}{\sinh (\m |y| +\m y_0 )}
+C^\pm (t)
.\ee
For $C^+ (t) =C^- (t)=C(t)$ the limits $y=0$ from both sides of the trajectory of the brane are the 
same.

From the fact that in (\ref{17}) the coefficient at $dydt$ is zero it follows that
\be
\l{32}
\dot{T} =\f{\dot{R}\,R'}{\m^4 R^4 \,T'}= \f{\dot{R}\,R'}{\m^2 R^2 (\e (y)E)}.
\ee
Writing (\ref{26}) as
\be
\l{27}
T(y,t)=-\f{ R' (y,t)/\e (y)}{\m^2 R(y,t) E} +C(t)
\ee
and taking the time derivative, we obtain
\be
\l{28}
\dot{T}= \f{\dot{R} R'}{\m^2 R^2 (\e (y) E )}-\f{1}{\m^2 R}\f{d}{dt}
{\left(\f{R' \,\e (y)}{E}\right)}
+\dot{C}
\ee
The first term in the rhs of (\ref{28}) written as
$$
\f{\dot{R} R'}{\m^2 R^2 (\e (y) E )}=\f{\dot{R}R'}{(\m^2 R^2 )^2 T'}
$$
 is the same as in (\ref{116}), which was obtained assuming that in (\ref{15}) $\p =0$.
Remarkably, substituting explicit expressions (\ref{25}) for $R$ and $R'$, we find that the second 
term on the rhs of (\ref{28}) is independent of $y$
\be
\l{29}
\f{1}{\m^2 R}\f{d}{dt}{\left(\f{R' \,\e (y)}{ E}\right)}=\f{\eta \dot{y}_0 (t)}{\dot{R}_b (t)}
.\ee
Choosing
\be
\l{30}
\dot{C}=\eta \f{\dot{y}_0}{ \dot{R}_b}
,\ee
 we obtain  $\dot{T}$ as in (\ref{32}). In the limit $y=0$ we have
\be
\l{31}
\dot{T}(y,t)|_{y\rightarrow 0}=-\eta\f{\r +\m }{\m^2 R_b}=\dot{T}_b (t)
.\ee
In the radiation-dominated period, conservation equation for the energy density on the brane is
$$
\dot{\r}=-4\r \dot{R}_b /R_b.
$$
Using this equation, we have
$$
\dot{y}_0
=-\f{4\r}{H},\qquad \dot{T}_b =-\eta\f{\r +\m }{\m^2 R_b}
.$$
From (\ref{21}), using the conservation equation for $\r$, we obtain
\be
\l{145}
\dot{R}(y,t)=
\f{\dot{R}_b}{2}\left[e^{\m |y|}\f{3\r}{\m}-e^{-\m
|y|}\left(\f{3\r}{\m}-2\right)\right]
.\ee
Jacobian of transformation from $T,\,R$ to $t,\, y$ is
$$
J= \dot{T}R'-\dot{R}T' =\e\eta\f{\dot{R}}{\dot{R}_b}.
$$
Although both $\dot{R}$ and $\dot{R}_b$ are non-zero, the expressions for
$\dot{T}$ and $T'$ contain $R(y,t)$ in denominator. $R(y,t)$ is zero for $e^{2\m |y|} =(\r +2\m
)/\r $,
 and transformation from $(T,\,R)$ to $(t,\, y)$
is valid for $e^{2\m |y|}< 1+2\m/\r$.
In the region $e^{2\m |y|}> 1+2\m/\r \,\,R(y,t)$ can be defined as
$$
R(y,t) =\f{R_b (t)}{2}\left[e^{\m |y|}\f{\r}{\m}-e^{-\m |y|}\left(\f{\r}{\m}+2\right) \right]
,$$
which is a positive and increasing function $y$.

I would like to thank I. Tyutin and B. Voronov for helpful discussion.
%%%%%%%%%%%%%%%%%%%%%%%%%%%%%%%%%%%%%%%%%%%%%%%%%%%%%%%%%%%%%%%%%%%%%%%%%%%%%%%%%%%%%%%%%%%%%%%%%%%%%%%%%%%%%$

\end{document}